\documentclass[preprintnumbers,showpacs,amsmath,amssymb,floatfix,prd,twocolumn,superscriptaddress,nofootinbib]{revtex4}
\usepackage{graphicx}
\usepackage{epsfig}
\usepackage{amsmath}
\usepackage{bm}
\usepackage{amsfonts}
\usepackage{epstopdf}
\usepackage{setspace}
\usepackage{neuralnetwork}
\usepackage{hyperref}

\begin{document}

\title{ Artificial Neural Network for Constructing 
	Type Ia Supernovae \\Spectrum Evolution Model}

\author{Qiao-Bin Cheng}
\affiliation{Shanghai United Center for Astrophysics (SUCA), \\ Shanghai Normal University,
	100 Guilin Road, Shanghai 200234, P.R.China}

\author{Chao-Jun Feng}
\email{fengcj@shnu.edu.cn} 
\affiliation{Shanghai United Center for Astrophysics (SUCA), \\ Shanghai Normal University,
    100 Guilin Road, Shanghai 200234, P.R.China}

\author{Xiang-Hua Zhai}
\email{zhaixh@shnu.edu.cn} 
\affiliation{Shanghai United Center for Astrophysics (SUCA), \\ Shanghai Normal University,
	100 Guilin Road, Shanghai 200234, P.R.China}

\author{Xin-Zhou Li}
\email{kychz@shnu.edu.cn} \affiliation{Shanghai United Center for Astrophysics (SUCA),  \\ Shanghai Normal University,
    100 Guilin Road, Shanghai 200234, P.R.China}

\begin{abstract}
We construct and train an artificial neural network  called  the back-propagation neural network  to describe the evolution of the type Ia supernova  spectrum by using the data from the CfA Supernova Program. This network method has many attractive features, and one of them is that the constructed model is differentiable. Benefitting from this, we calculate the absorption velocity and its variation. The model we constructed can well describe not only the spectrum of SNe Ia with wavelength range from $3500\AA$ to $8000\AA$, but also the light-curve evolution with phase time from $-15$ to $50$ with different colors. Moreover, the number of parameters needed during the training process is much less than the usual methods.
\end{abstract}

\maketitle


\section{Introduction}
Type Ia supernova (SN Ia) is regarded as the standard candle to measure the distance on cosmological scales, since all of them have almost the same intrinsic brightness. By using the SN Ia distance indicator, the accelerating expansion of the universe was discovered in 1998 \cite{Riess:1998cb,Perlmutter:1998np}. The standard model of SN Ia involves the thermonuclear disruption of a carbon-oxygen white dwarf star as it approaches the Chandrasekhar mass, and most type Ia supernovae are very similar in their photometric and spectroscopic properties. The SN Ia may be generated through the merging of components in close binaries \cite{Bogomazov}. A correlation between the peak luminosity and the shape of the early light curve was also found, with brighter objects having a lower rate of decline than dimmer ones \cite{Phillips:1993ng}, which is matched by a spectroscopic sequence.

The model of the spectroscopic sequence is  constructed by a training process, such as the SALT2 model\cite{Guy:2007dv}, in which the mean evolution of the spectral energy distribution (SED) sequence of SN Ia and its variation with color are modeled as a functional form, see Eq.(1) in\cite{Guy:2007dv}. During the training process of SALT2, one ends up with  more than $3000$ parameters\cite{Guy:2007dv} to fit, due to the obvious non-linearities of the SED.

In the past decades, among the various machine intelligence procedures, artificial neural network (ANN) methods have been established as powerful techniques to solve a variety of real-world problems because of its excellent learning capacity \cite{McCulloch:1943,Minsky}. ANN is one of the popular areas of artificial intelligence research and also an abstract computational model based on the organizational structure of the human brain \cite{McCulloch:1943}. In this paper, we will construct and train an ANN called  the back-propagation neural network (BPNN) to describe the evolution of the SN Ia spectrum. The inputs of our ANN are the phase (or time), wavelength and also the color from the data that will be described later, while the output is just the corresponding flux of the SN Ia.  The SED we trained is almost the same as that in the SALT2 model. However, the employment of a neural architecture adds many attractive features:
\begin{itemize}
	\item One does't need to assume a functional form of the SED model during the ANN's training process.
	\item The SED via ANNs is differentiable, then it is easily used in any subsequent calculations, e.g. the calculation of the absorption velocity gradients of Si II $\lambda 6355$ line  of SN Ia in the following.
	\item The required number of model parameters is far less than the traditional methods.
	\item The method can be realized in hardware, using neuroprocessors, and hence offer the opportunity to handle the real-time analysis of the SN Ia's spectra.
	\item The method can also be efficiently implemented on parallel architectures, such as the data parallelism, in which the data set is split into small ones and each of them is feed to a processor.
\end{itemize}

Si II $\lambda 6355$ is one of the strongest features in optical/near-infrared spectra of SN Ia; the blueshift of its absorption minimum has often been used to diagnose the diversity among SN Ia \cite{Wang:2009fs}. Different subclass evolutions of Si II $\lambda 6355$  absorption velocity ($v_{abs}$) have been compared in Ref.\cite{Blondin:2012ha}, while the relation between its gradient and the parameter $\Delta m_{15}$ of SN Ia has been studied in \cite{Benetti:2004bk}. In this paper, we have studied the properties of $v_{abs}$ and its gradients by using the spectrum evolution model after ANN's training. The results are fully consistent with those in \cite{Branch:2006jd,Blondin:2005yx}.  

The structure of this paper is as follows. In Section \ref{sec:netConstruction}, we construct the neural network, and then train it in  Section \ref{sec:netTraining}.
The spectra data set we used for training  is described in Section \ref{sec:data}, while the training  results are presented
in Section \ref{sec:results}, and the $v_{abs}$ and its gradient are also computed in this section. In Section \ref{sec:abs} the relation between $v_{abs}$ and color is shown.
Finally,discussions and conclusions is given in Section \ref{sec:conclusion}.

\section{Network Construction}\label{sec:netConstruction}
In the following, the neural network will be constructed, which is called the back-propagation neural network. This kind of ANN has  already been used in astronomy and physics, for example, to classify the type of a supernova (e.g.Ia,  Ib,  II,  etc.) by using the neural network from the  multi-frequency observations of its light curve, see \cite{Graff:2013cla} and references therein. The structure (or topology) of the BPNN could be described as Fig.\ref{fig:net}:
\begin{figure}[hb]
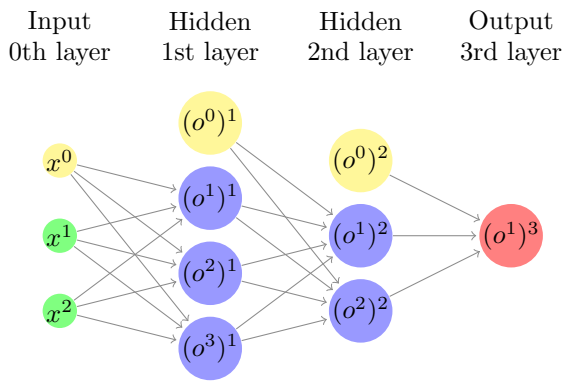

\begin{center}
	\begin{neuralnetwork}[nodespacing=10mm, layerspacing=20mm,
		maintitleheight=2.5em, layertitleheight=2.5em,
		height=4.5, toprow=false, nodesize=10pt, style={},
		title={}, titlestyle={}]
		\newcommand{\nodetextclear}[2]{}
		\newcommand{\nodetextx}[2]{$x^#2$}
		\newcommand{\nodetexty}[2]{$y_#2$}
		\newcommand{\nodetexto}[2]{$(o^#2)^#1$}
		\inputlayer[count=2, bias=true, title=Input\\0th layer , text=\nodetextx]
		\hiddenlayer[count=3, bias=true, title=Hidden\\1st layer, text=\nodetexto] \linklayers
		\hiddenlayer[count=2, bias=true, title=Hidden\\2nd layer, text=\nodetexto] \linklayers
		\outputlayer[count=1, title=Output\\3rd layer, text=\nodetexto] \linklayers
	\end{neuralnetwork}
	\caption{Typical structure of an ANN.}
	\label{fig:net}
\end{center}
\end{figure}
where a few hidden layers and a few neural neurons in each layer are presented for limited space. In general, there could be totally $L+2$ layers in the network, and we will use  $l=0,1,2,\cdots,L,L+1$ to denote them, where $l=0$ is called the input layer, $l=L+1$ the out put layer, and others are all called the hidden levels. 

Let $N^l$ denote the number of neurons at level $l$, with $N^0$ and $N^{L+1}$ being also called the dimensions of the input and the output, respectively. In this paper, we consider two cases with  $N^0=2$ and $3$, and $N^{L+1}=1$ for both. 

Between the $l$-th and the ($l$-1)-th level, we have the weights :
\begin{equation}
\mathbf{W}^l \equiv \left(W_{j}^i\right)^l  \in \mathcal{R}^{N^{L}\times (N^{L-1}+1) }\, 
\end{equation}
for $l\ge 1$, where $i=1,2,\cdots, N^l\,, j=0,1,2,\cdots, N^{l-1}$, and here we have included the bias vector in $j=0$, namely
\begin{equation}
\left(W_{0}^i\right)^l = (b^i)^l\,.
\end{equation}

By construction, the input of the $l$-th level is just the output of the ($l$-1)-th level plus the bias, so we set the  output of the $l$-th level $(o^i_m)^l$ as the following:
\begin{eqnarray}
(o^i_m)^0 &=& (n^i_m)^0 = X^i_m\,, \quad l=0\,,	\\
(n^i_m)^l &=&  \sum_{j=0}^{N^{l-1}}\left(W^i_{j}\right)^l (o^j_m)^{l-1}\,,\quad l\geq 1 \,,\\
(o^i_m)^l &=& g_l((n^i_m)^l)\,,\quad l\geq 1 \,,
\end{eqnarray}
where $i= 1,2,\cdots, N^l$, $m=1,2,\cdots,M$ and $M$ denotes the total number of the samples. Here, the function $g_l(x)$ is  the activation function, which is often taken as a sigmoid function or the tangent hyperbolic function
\begin{equation}
g(x) = \left\{ \begin{array}{ll}
\frac{1}{1+e^{-x}} \, & \textrm{for sigmoid function}\,, \\
\textrm{tanh(x)} \,& \textrm{for tanh(x)}\,,
\end{array}\right.
\end{equation}
between the hidden layers, i.e. $l\leq L$. Straightforward, we also have:
\begin{equation}
\dot g=\frac{dg}{dx}= \left\{ \begin{array}{ll}
g(1-g) \,& \textrm{for sigmoid function} \,,\\
1-g^2 \,&  \textrm{for tanh(x)} \,.
\end{array}\right.
\end{equation}
For the output layer, we take the activation function  as a linear function $g(x) = x, \dot g=1$.

\section{Network Training} \label{sec:netTraining}

To train the network, one needs to minimize a cost function after feeding training samples. The cost function is also called the error function, which describes  the error between the output and training samples. In our cases, the cost function is given by $E = \mathbf{ e^T\mathcal{C}^{-1} e}/2$, or more specifically:
\begin{equation}\label{SSE}
	E = \frac{1}{2}\bigg(\vec{F}^{\text{obs}} - \vec{F}^{\text{ANN}}\bigg)^T
	\mathcal{C}^{-1}
	\bigg(\vec{F}^{\text{obs}} - \vec{F}^{\text{ANN}}\bigg) \,,
\end{equation}
where $\mathcal{C}$ is the covariance matrix of the observational flux $\vec{F}^{\text{obs}}$
and the flux $\vec{F}^{\text{ANN}}$ is  the output of BPNN for a supernova at the redshift $z$ that depends on the phase $p \equiv(t-t^B_{\text{max}})/(1+z)$, the wavelength $\lambda$ and maybe the color $C$. 
Furthermore, the error for a given sample $m$ in Eq.(\ref{SSE}) reads:
\begin{equation}\label{eq:error}
e_m  = F_m^{\text{obs}} - F_m^{\text{ANN}} \,.
\end{equation}

In the following, we will perform the Levenberg-Marquardt (LM) algorithm to train the BPNN. In each step $s$, the following weighted normal equations will be solved
\begin{equation}
\mathbf{\left(J^T\mathcal{C}^{-1}J + \mu I \right)}\bigg(W_{s+1}-W_s\bigg)=-\mathbf{J^T\mathcal{C}^{-1}e}\,,
\end{equation}
to update each weight between the layers of the network for next step $s+1$. Here $\mathbf{I}$ is the identity matrix, $\mu$ is the combination coefficient that could be changed during the training procedure. The Jacobi matrix $\mathbf{J} \in \mathcal{R}^{M\times N}$ is defined as
\begin{equation}
\begin{pmatrix} 
\frac{\partial e_1}{\partial W_1} & \frac{\partial e_1}{\partial W_2} &\cdots& \frac{\partial e_1}{\partial W_N}\\ 
\frac{\partial e_2}{\partial W_1} & \frac{\partial e_2}{\partial W_2} &\cdots& \frac{\partial e_2}{\partial W_N}\\ 
\cdots&\cdots&\cdots&\cdots\\
\frac{\partial e_M}{\partial W_1} & \frac{\partial e_M}{\partial W_2} &\cdots& \frac{\partial e_M}{\partial W_N}\\ 
\end{pmatrix}\,,
\end{equation}
where $N = \sum_{l=1}^{L+1} N^{l}(N^{l-1}+1)$ is the total number of  weights ( including bias). 

For a given sample $m$ we also have the following recurrence relation
\begin{eqnarray}
		\frac{\partial e_m}{\mathbf{W}^l} &=& \frac{\partial{(F^{obs}_m -F^{ANN}_m)}}{\mathbf{W}^l} = \frac{-\partial(o_m^{L+1})}{\mathbf{W}^l} \,,\\
		\frac{\partial (o_m^{L+1})}{\mathbf{W}^l} &=& \delta^l (\mathbf{o}_m^{l-1})^T \,,\\
		\delta^{L+1} &=& 1 \,,\\
		\delta^{l} &=& \dot G^l(\mathbf{n}_m^l) (\mathbf{\bar W}^l)^T \delta^{l+1} \,,\quad l\leq L\,,
\end{eqnarray}
where $(\mathbf{\bar W}^l)$ are the weights not including the bias vector, i.e. $(\bar W^i_j)^l = (\bar W^i_j)^l, j\neq 0$, and 
\begin{equation}
\dot G^l(\mathbf{n}_m^l) = 
\begin{pmatrix}
	\dot g^l((n_m^1)^l) &0&\cdots& 0 \\
	0&\dot g^l((n_m^2)^l) &\cdots& 0 \\
	\cdots&\cdots&\cdots &\cdots \\
	0 &0&\cdots&\dot g^l((n_m^{N^l})^l) \\
\end{pmatrix}\,.
\end{equation}
In summary, by using LM algorithm, the update rule of weights can be presented as
\begin{equation}\label{eq:updateWeights}
W_{s+1}=W_s-\mathbf{\left(J^T\mathcal{C}^{-1}J + \mu I \right)^{-1}}\mathbf{J^T\mathcal{C}^{-1}e}\,.
\end{equation}
Actually, we also have tried the famous RPROP  algorithm\cite{rprop} to train BPNN, but it turns out that the LM one is much better, especially when the input dimension is higher than 2. 

\section{Data Analysis}\label{sec:data}

The data used in this paper are obtained through the CfA Supernova Program, and there are 2603 spectra of 462 low-z SNIa in total, see Refs.\citep{Blondin:2012ha,Matheson:2008pa,Branch:2003hk,Jha:1999sm,Krisciunas:2011sn,Li:2003wja,Foley:2009wk,Hicken:2007ap}. For the training, the recalibration (or re-normalization) of spectra from the corresponding supernova needs the knowledge of $t_{max}^B$, i.e. the time when the flux has a maximum luminosity, so that spectra without the information of $t_{max}^B$ will be disregarded. We only focus on  $-15<p < 50$, which is the most important phase range when dealing with the spectra. Beyond this range, we have either few spectra or slowly varying ones. The spectra of SN Ia with color $C>0.8$ and those without a spectrum in range $-10<p<15$ are ignored, since the supernova has strong extinctions for large color.  Here, the color of a SN Ia could be defined through the spectra: 
\begin{equation}\label{eq:calcColor}
C = 2.5\log\Bigg[
\frac{\int_{\lambda}\lambda T_V(\lambda)S_{SN}(\lambda)d\lambda}{\int_{\lambda}\lambda T_B(\lambda)S_{SN}(\lambda)d\lambda}
\Bigg]\,,
\end{equation}
where $S_{SN}(\lambda)$ is the de-redshift spectrum  nearest $t_{max}^B$, and 
$T_B(\lambda)$ and $T_V(\lambda)$ are the effective instrument transmissions in photometric band B and V, respectively. The color defined from spectroscopy in the equation above  is almost the same properties as that defined from the photometry, see Ref.\cite{Matheson:2008pa}.

\begin{figure}[h]
	\begin{center}
		\includegraphics[width=0.45\textwidth,angle=0]{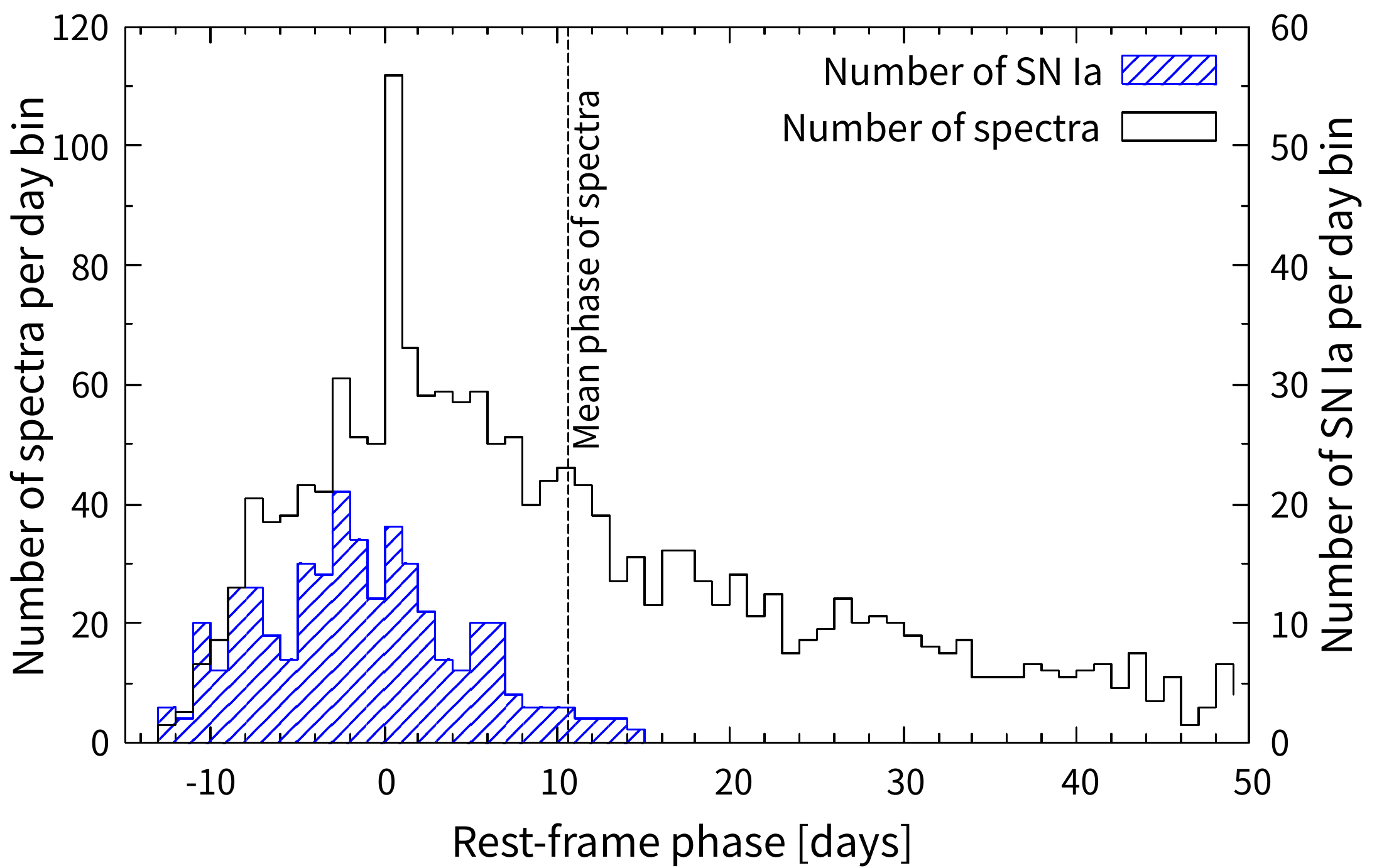}
		\caption{Number distributions of the spectra and SN Ia.}
		\label{fig:numOfSpe}
	\end{center}
	\begin{center}
		\includegraphics[width=0.45\textwidth,angle=0]{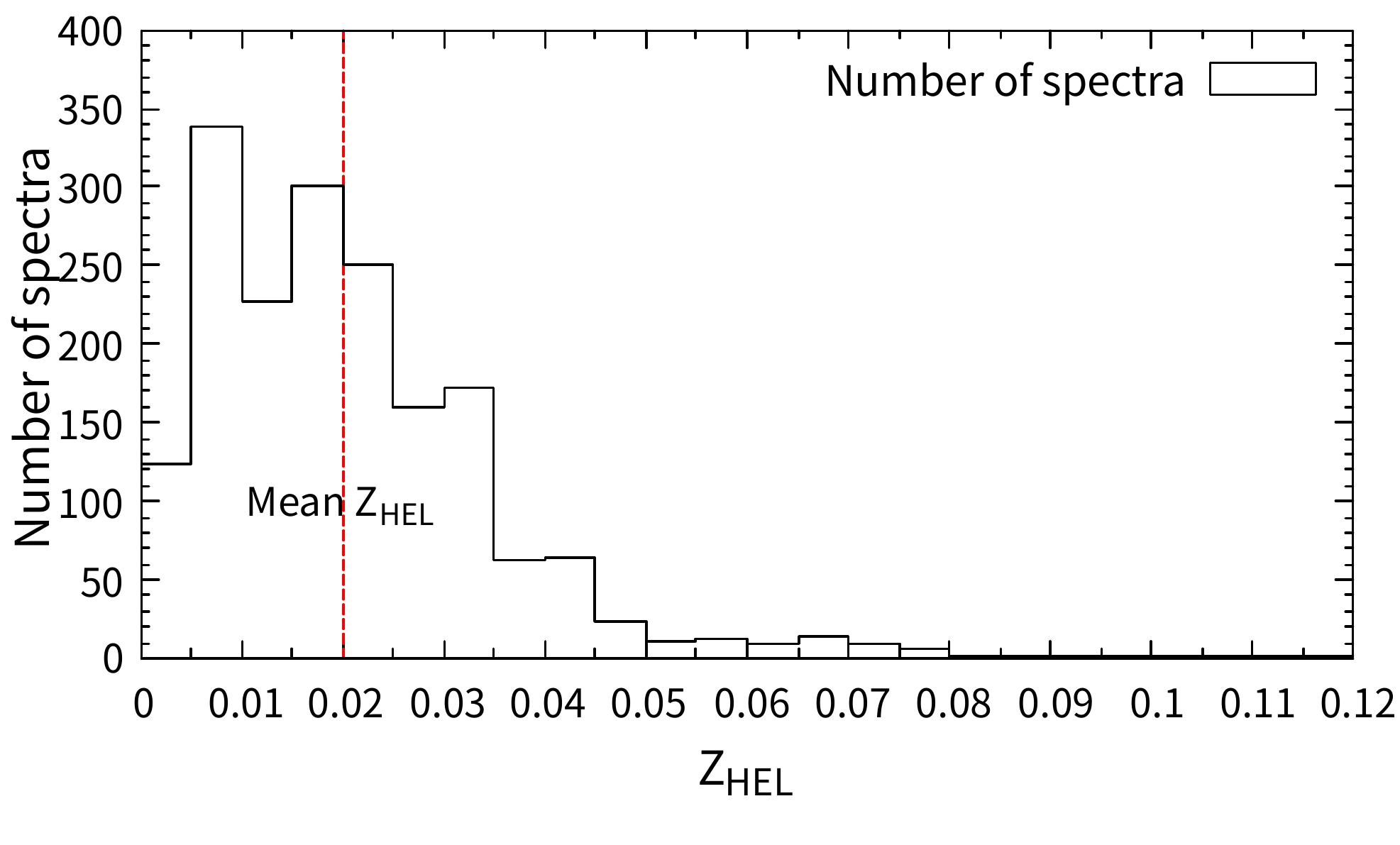}
		\caption{Redshift distribution of the spectra.}
		\label{fig:redshift}
	\end{center}
\end{figure}

Finally, there are 1787 spectra of 238 SN Ia (about 4,600 thousand data points) will be left for the BPNN training. We have shown the number and redshift distributions of these samples in Fig.\ref{fig:numOfSpe} and Fig.\ref{fig:redshift}. From the number distribution of the spectra in Fig.\ref{fig:numOfSpe} (unfilled histogram), we find that many spectra were obtained near the max date $t_{max}^B$ ($p=0$), but the average phase is about $\bar p \sim 10.5$ because the distribution has a long tail. The height of each column of the filled histogram in Fig.\ref{fig:numOfSpe} indicates the number of SN Ia whose earliest spectrum is obtained at the phase $p$. It shows that more than half of the supernovae have the spectra earlier than the max date. From Fig.\ref{fig:redshift}, one can see that almost all the SN Ia are below $z<0.05$, and the average value is $\bar z\sim 0.02$.

\begin{figure}[h]
	\begin{center}
		\includegraphics[width=0.45\textwidth,angle=0]{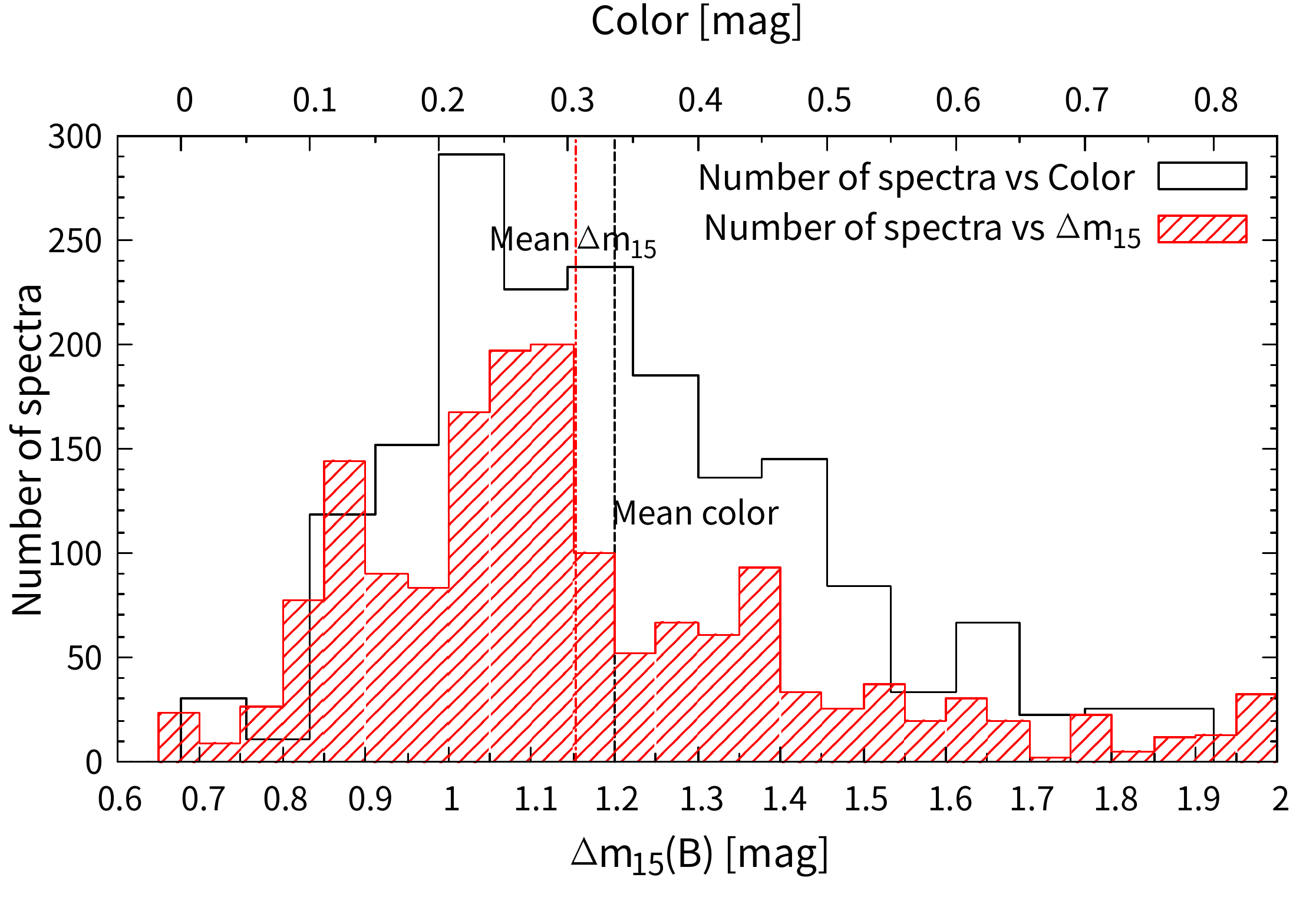}
		\caption{Parameters distribution of the spectra.}
		\label{fig:dm15AndColor}
	\end{center}
\end{figure}

The $\Delta m_{15}$ and color $C$ distributions of the spectra are also plotted in Fig.\ref{fig:dm15AndColor}. The filled histogram tells us the average value of $\Delta m_{15}$ is about $\bar{\Delta m_{15}} \sim 1.15$ mag, while the unfilled one shows the average value of $\bar C \sim 0.34$ mag.

To make best use of the spectroscopic data, one needs to recalibrate or re-normalize the data by using the corresponding SN Ia photometry, since in general the photometry  is much better calibrated and uniform than its spectra. The recalibration is performed by multiplying each input spectrum by the factor:
\begin{equation}
f_{SN}^p=\frac{\bar F^{obs}_R(p)}{\int_\lambda \lambda T_R(\lambda)S_{SN}(\lambda)_pd\lambda}
\end{equation}
where $\bar F^{obs}_R(p)$ is the average value of the flux  in the R band filter.

\section{Training Results}\label{sec:results}

The output flux of BPNN $\vec{F}^{ANN}$ in Eq.(\ref{SSE}) could either depend on two inputs, i.e. phase $p$ and wavelength $\lambda$, or three inputs, phase $p$, wavelength $\lambda$ and color $C$. In the following, the networks of these two cases will be trained. For simplicity, we assume each hidden level of the network has the same number of neurons in both cases. 

\subsection{Case I: $\vec{F}^{ANN} = \vec{F}^{ANN}(p,\lambda)$}

In this case, the network will be trained by using the spectra in the  wavelength range between $3500 \AA$ and $8000\AA$ with a bin interval of $10\AA$. In each interval, we average the flux weighted by its variation. There are about 57,000 data points in total.

\begin{figure}[h]
	\begin{center}
		\includegraphics[width=0.45\textwidth,angle=0]{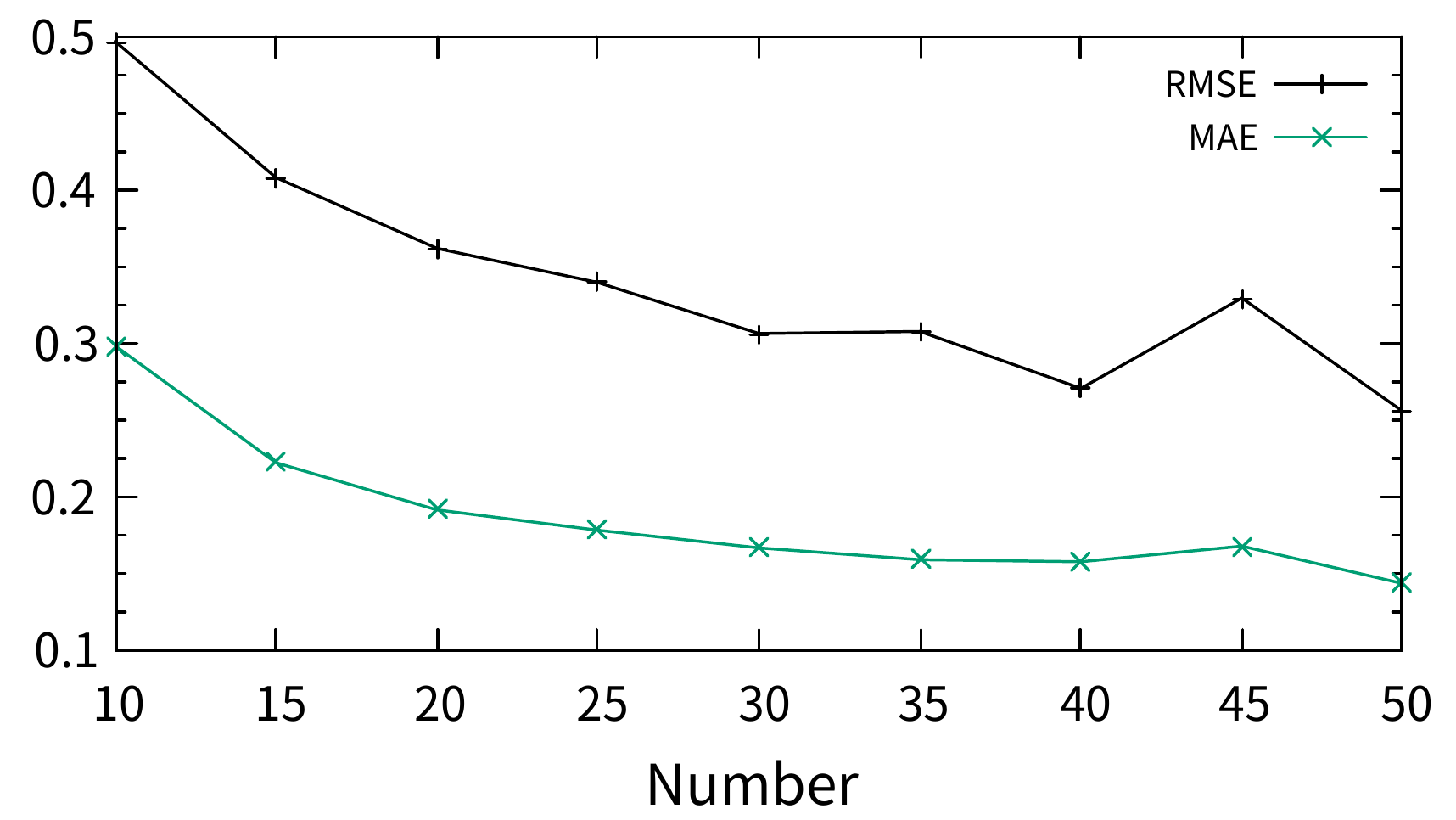}
		\caption{The RMSE and MAE vs the neuron number in the hidden level for the BPNN with only one hidden level.}
		\label{fig:MSEMAE}
	\end{center}
\end{figure}

\begin{figure}[!h]
	\begin{center}
		\includegraphics[width=0.5\textwidth,angle=0]{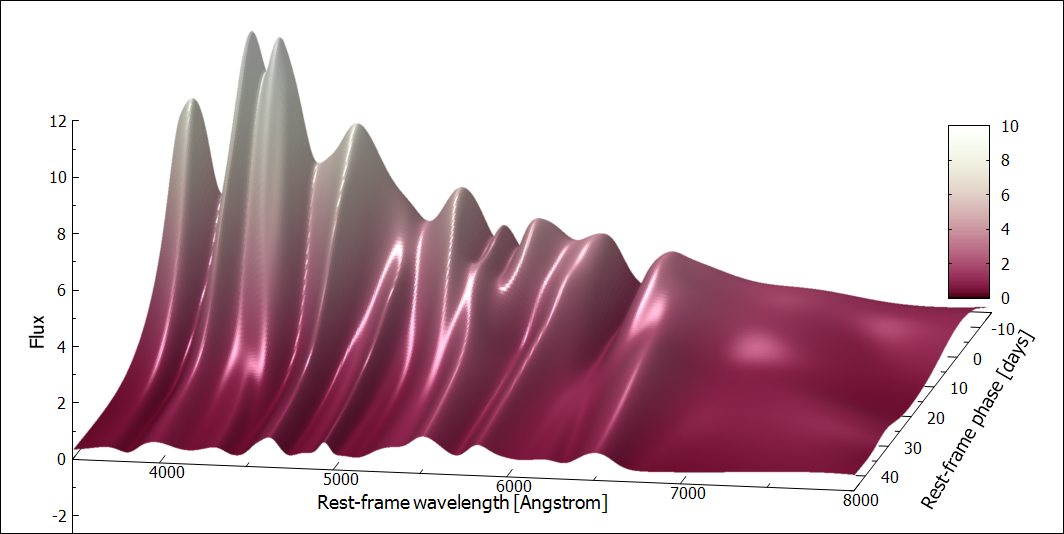}
		\caption{3D vision: A best-trained spectrum evolution model. }\label{fig:3dtemplate}
	\end{center}
\end{figure}

\begin{figure}[!h]
	\begin{center}
		\includegraphics[width=0.5\textwidth,angle=0]{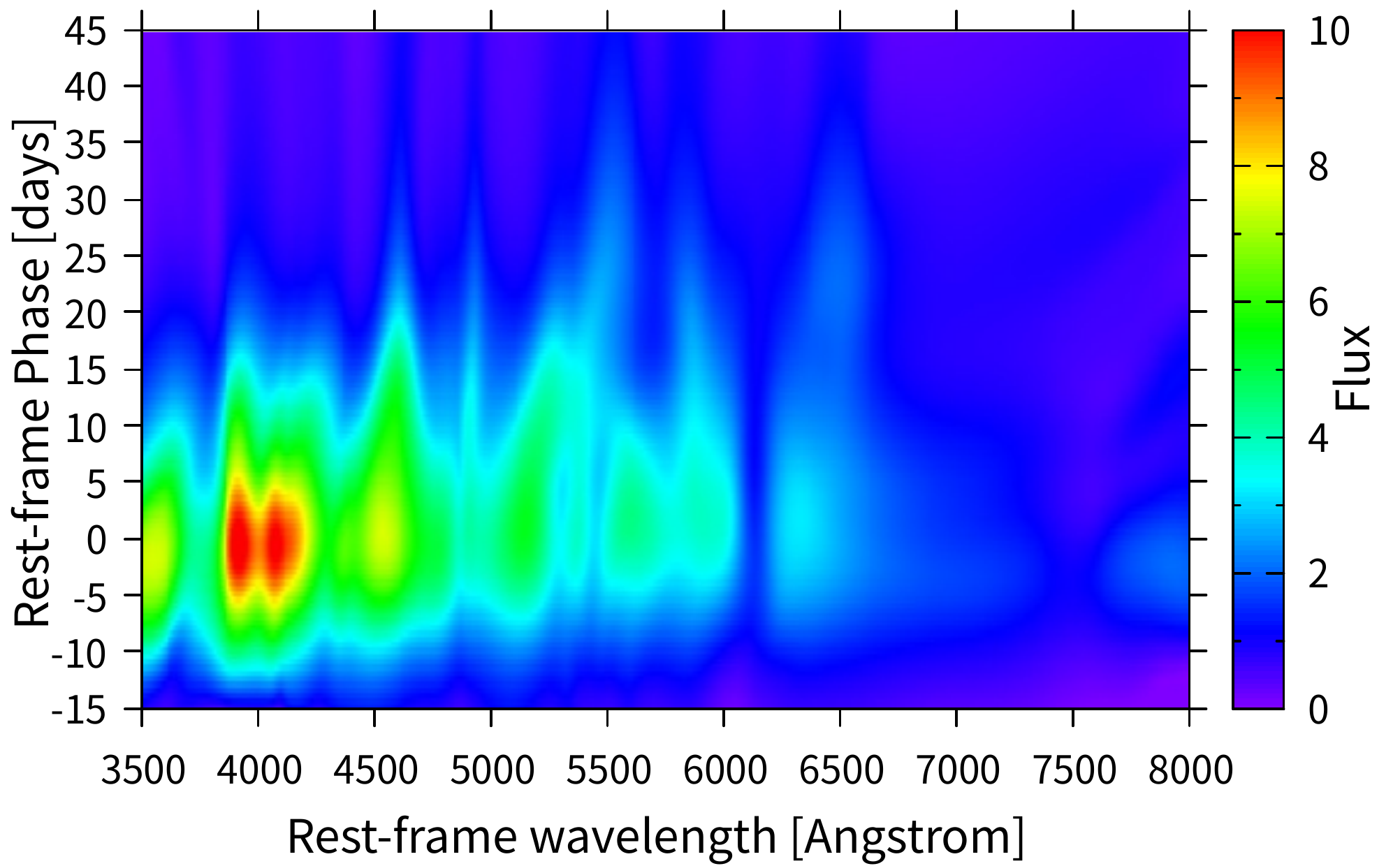}
		\caption{Contour: A best-trained spectrum evolution model. }\label{fig:2dtemplate}
	\end{center}
\end{figure}

\begin{figure}[!h]
	\begin{center}
		\includegraphics[width=0.4\textwidth,angle=0]{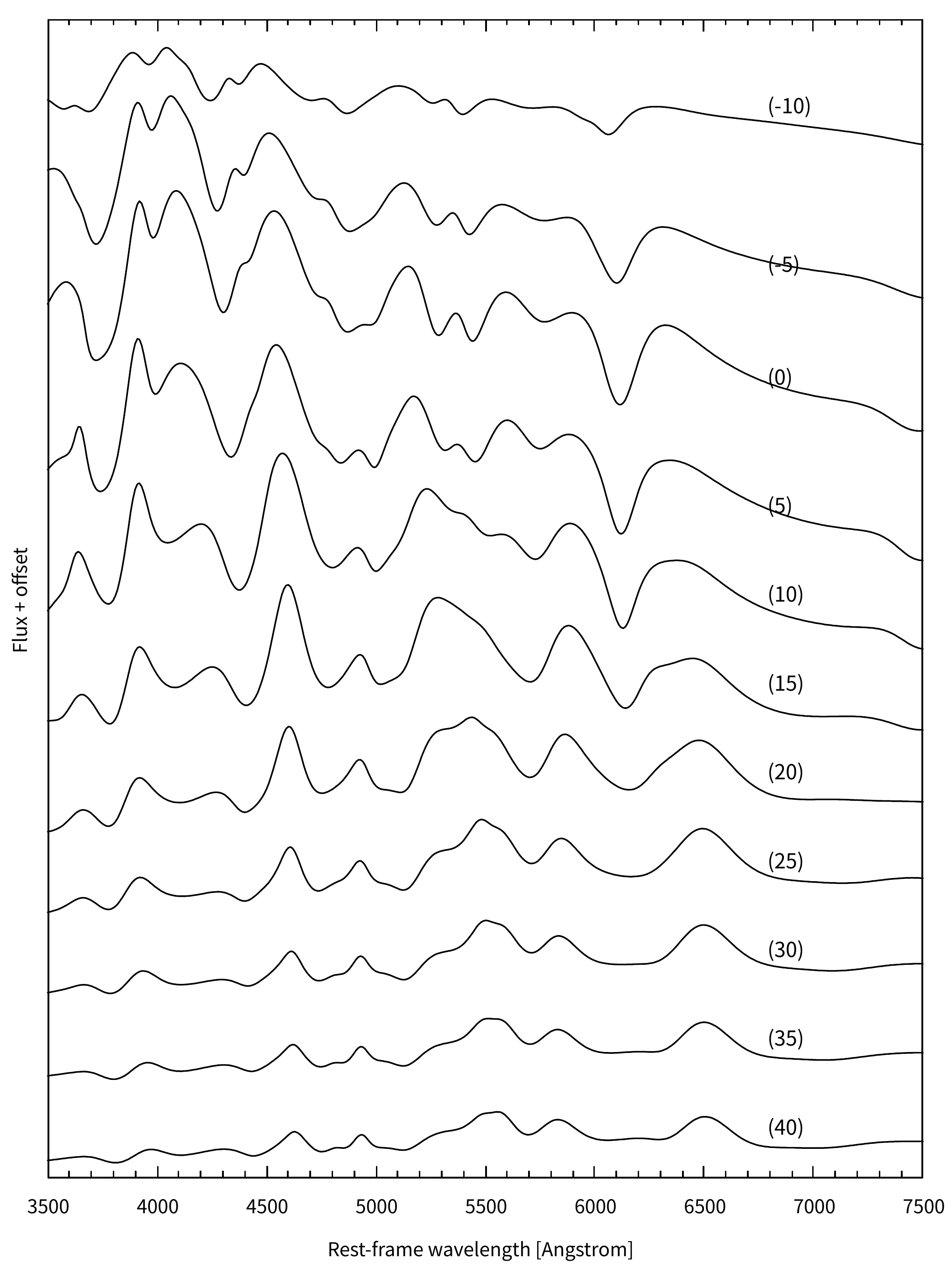}
		\caption{Evolution of spectrum with phase, where the number in parentheses denotes the phase.}
		\label{fig:evo}
	\end{center}
\end{figure}

Fig.\ref{fig:MSEMAE} indicates the variation of the root mean squared error (RMSE) and the mean absolute error (MAE)  vs the neuron number, where the BPNN only has one hidden level. One can clearly see that both RMSE and MAE decrease with the increasing of the neuron number. We have also compared different structure of network with three or four hidden levels, see Tab.\ref{tab:meanResuult}.  
\begin{table}[!hbp]
	\begin{tabular}{rccc}
		\hline
		\hline
		Structure & $\#$  & RMSE & MAE \\
		\hline
		2-10-10-1       &   151  & 0.4964 &0.2981\\
		2-15-15-1       &   301 & 0.4077 &0.2220\\
		2-20-20-1       &   501  & 0.3549 &0.1871\\
		\hline
		2-10-10-10-1    &   261  & 0.4063 &0.2206\\
		2-15-15-15-1    &   541  & 0.3491 &0.1866\\
		2-20-20-20-1    &   921  & 0.3059 &0.1690\\
		\hline
		2-10-10-10-10-1 &   371  & 0.3769 &0.1988\\
		2-15-15-15-15-1 &   781  & 0.3016 &0.1705\\
		2-20-20-20-20-1 &   1341 & 0.2780 &0.1665\\ 
		\hline
	\end{tabular}
	\caption{Results in the Case I with different network structure. The first column is the topology of the network, the second column is the total number of the weights (including the bias), and the third and forth columns are the final RMSE and MAE respectively. Note that, the values of RMSE and MAE are already divided by the number of data points. }\label{tab:meanResuult}
\end{table}
The total number of parameters including weights and bias in the training network is less than $1000$ except the last one, which has the most complicated structure. However,  during the training process of SALT2, one usually ends up with  more than $3000$ parameters\cite{Guy:2007dv} to fit. To illustrate the ability of the BPNN method, a best-trained spectrum evolution model is presented in Figs.\ref{fig:3dtemplate}-\ref{fig:evo}. 

\subsection{Case II: $\vec{F}^{ANN} = \vec{F}^{ANN}(p,\lambda, C)$}\label{sec:results:ac}

It is well known that the K-correction of light-curve mainly depends on spectral color\cite{Kim:1995qj,Nugent:2002si}, so in this case, the color will be included as one  input of the network and the full set of  spectra data will be used instead of the bin one.

\begin{figure}[h]
 	\begin{center}
 		\includegraphics[width=0.5\textwidth,angle=0]{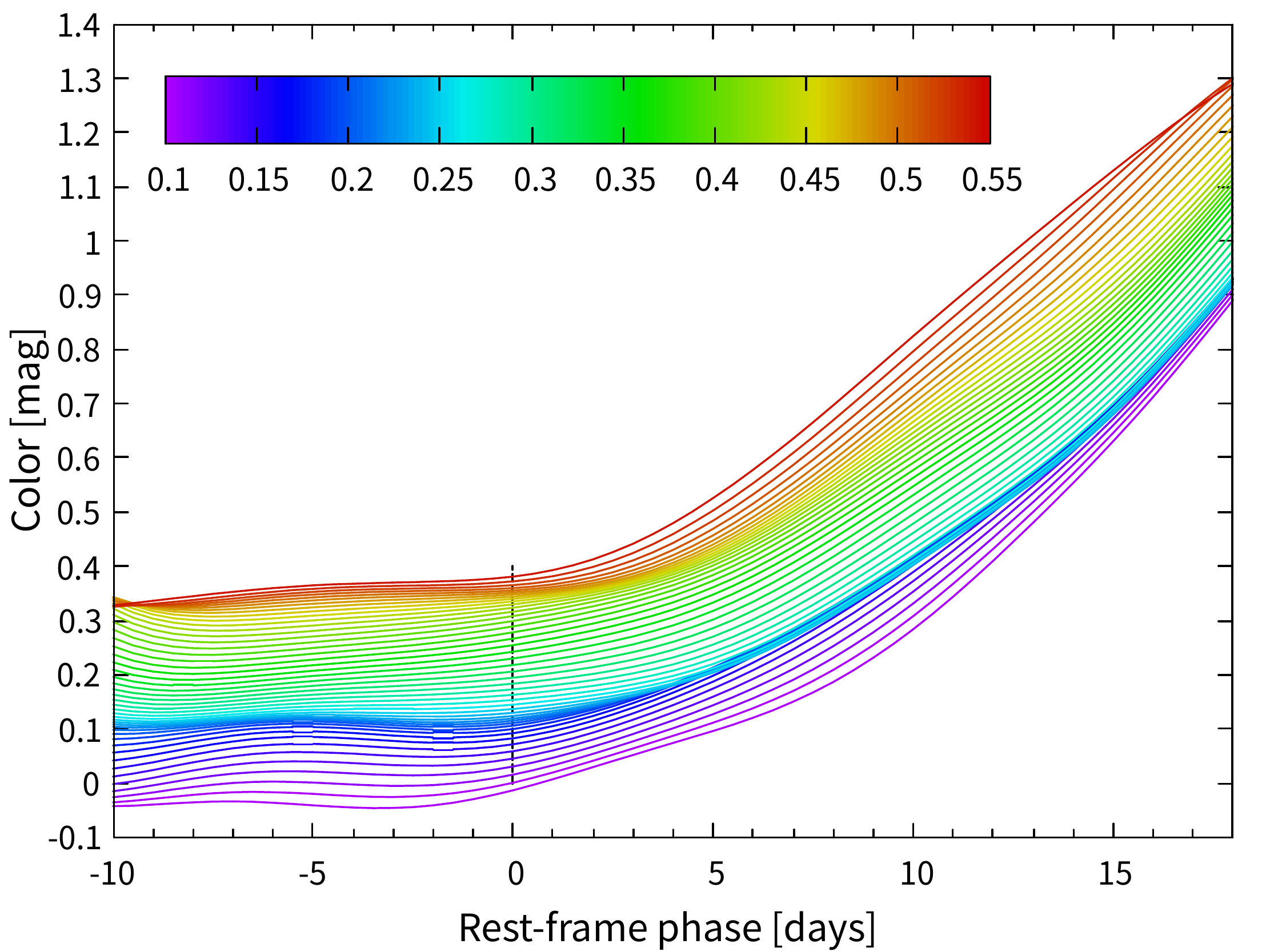}
 		\caption{Color vs phase.}
 		\label{fig:colorWithPhase}
 	\end{center}
 \end{figure}

\begin{figure}[!hbp]
\centering
\includegraphics[width=0.45\textwidth,height=1.5in]{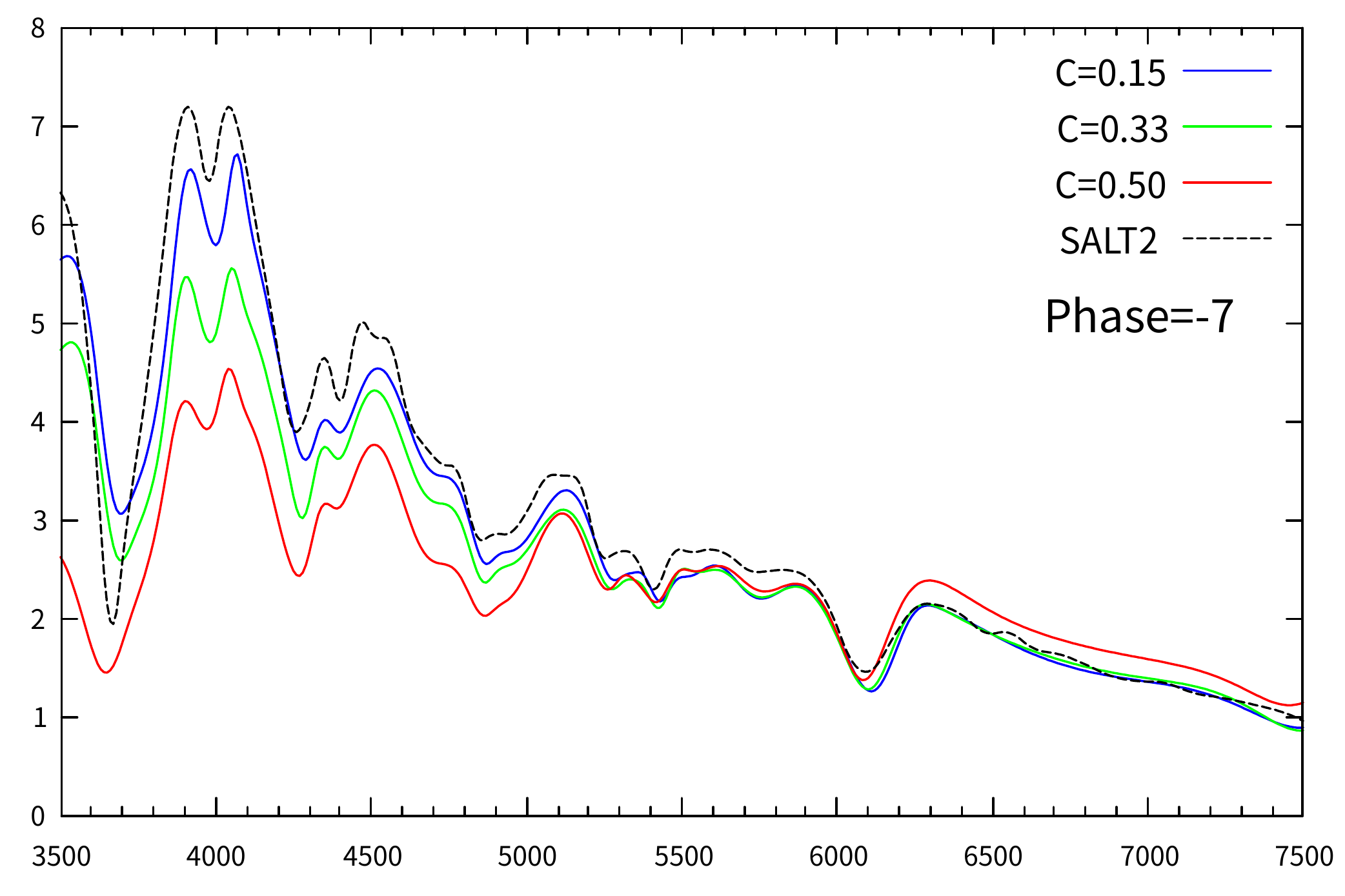}
\includegraphics[width=0.45\textwidth,height=1.5in]{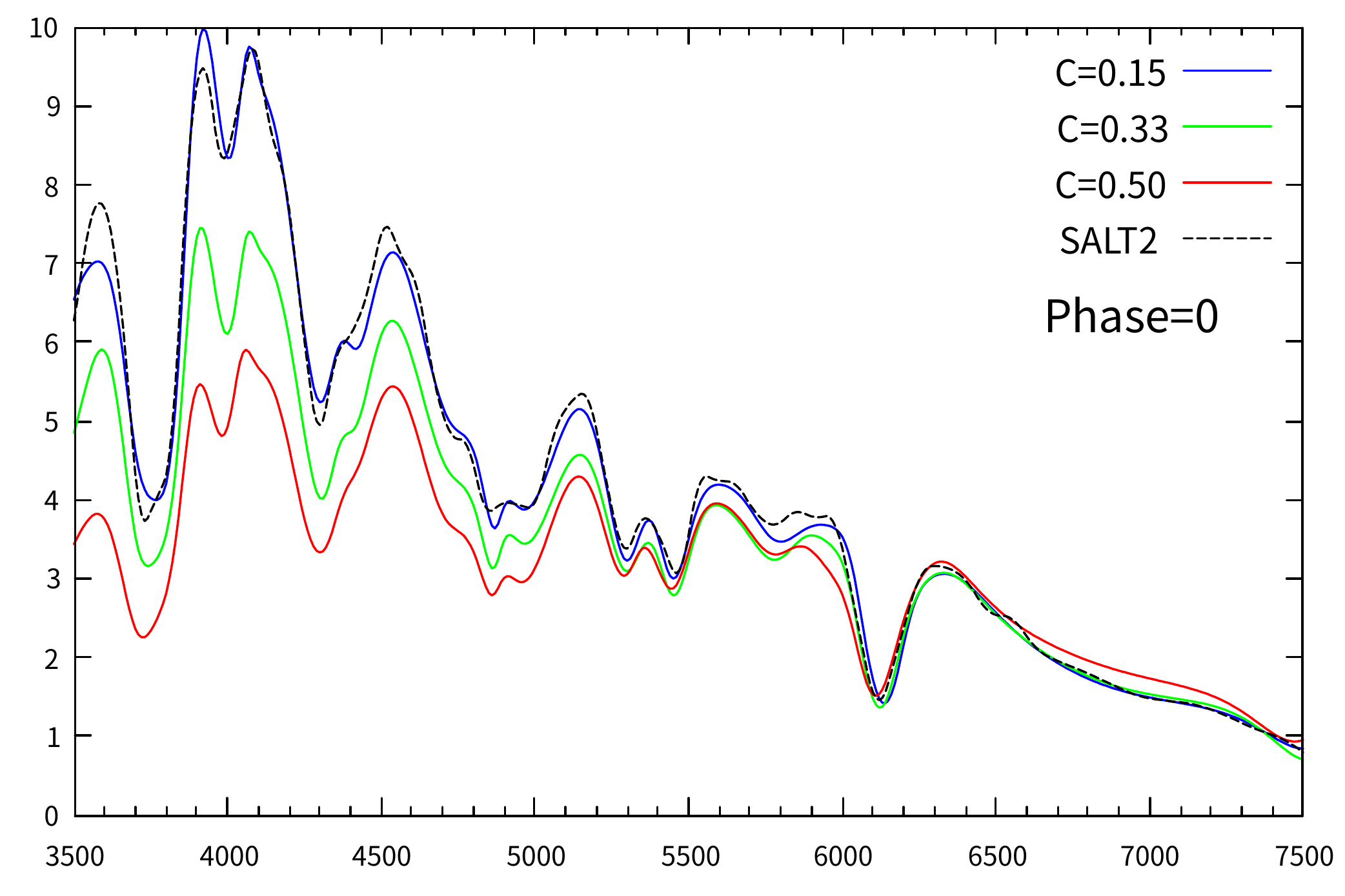}
\includegraphics[width=0.45\textwidth,height=1.5in]{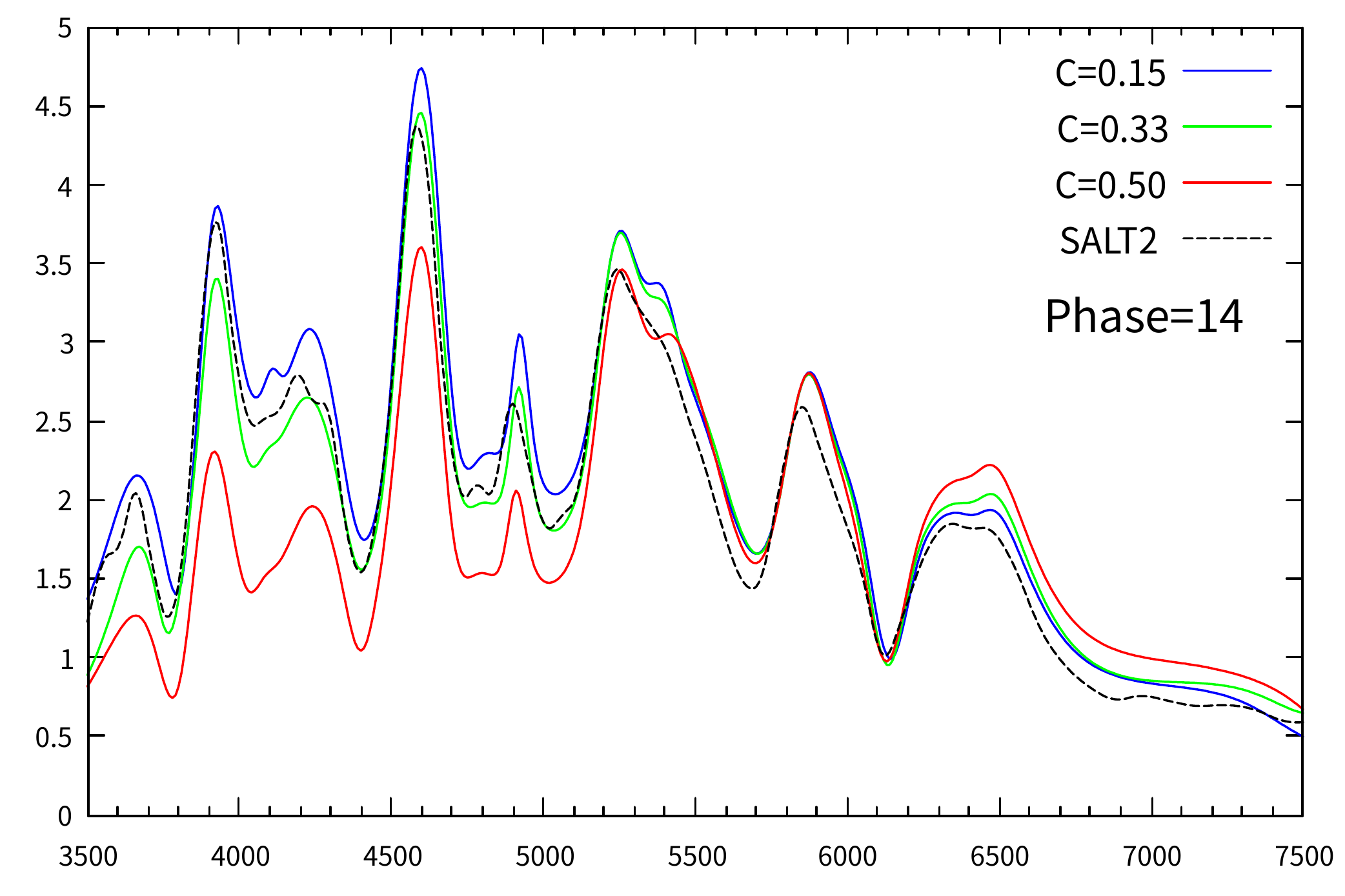}
\includegraphics[width=0.45\textwidth,height=1.5in]{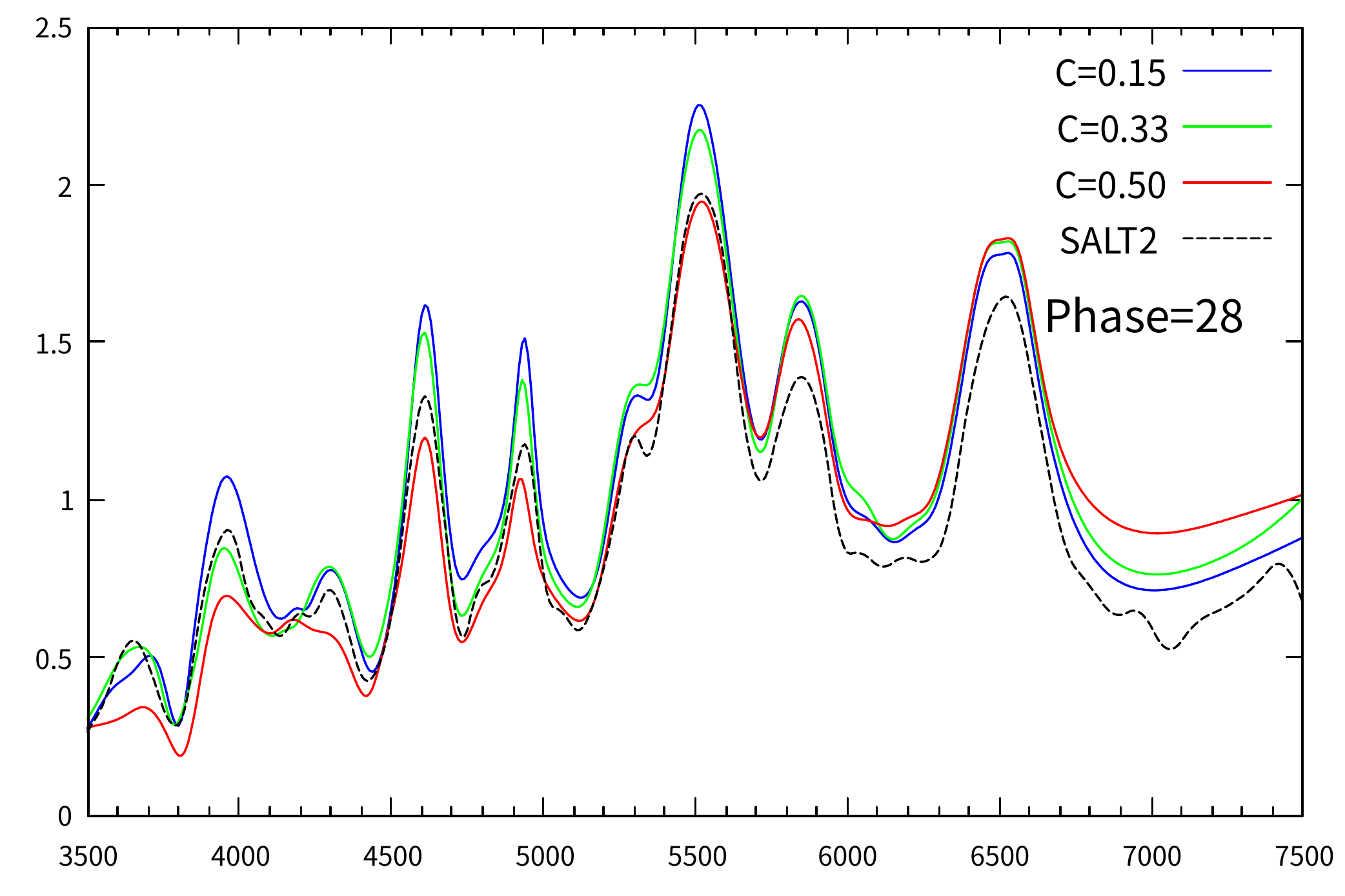}
\caption{Comparison of different color spectra. The black dash line is SALT2 template spectra.}\label{fig:compare}
\end{figure}

After training, we find the behavior of the RMSE and MAE is the same as that in the Case I. It is interesting that the color is almost a constant before the maximum luminosity day, after which it is proportional to the phase time, see Fig.\ref{fig:colorWithPhase}. 
We also compared the trained spectra with those in SALT2 at different phases, see Fig\ref{fig:compare}. One can see that the difference between them is very small at long wavelength, whereas  at the short wavelength, the difference mainly depends on the color value; when $C\sim0.15$, the model we trained is almost the same as that in SALT2. 

\section{Application: Absorption Velocity} \label{sec:abs}

As we stated in the introduction section, one of the attractive features of applying the ANN method is that the output of the network is differentiable. One can benefit from this property to perform the consequence calculation, e.g. to calculate the absorption velocity gradients of Si II $\lambda 6355$ line of SN Ia.

\begin{figure}[!hbp]
	\includegraphics[width=0.45\textwidth,height=2.3in]{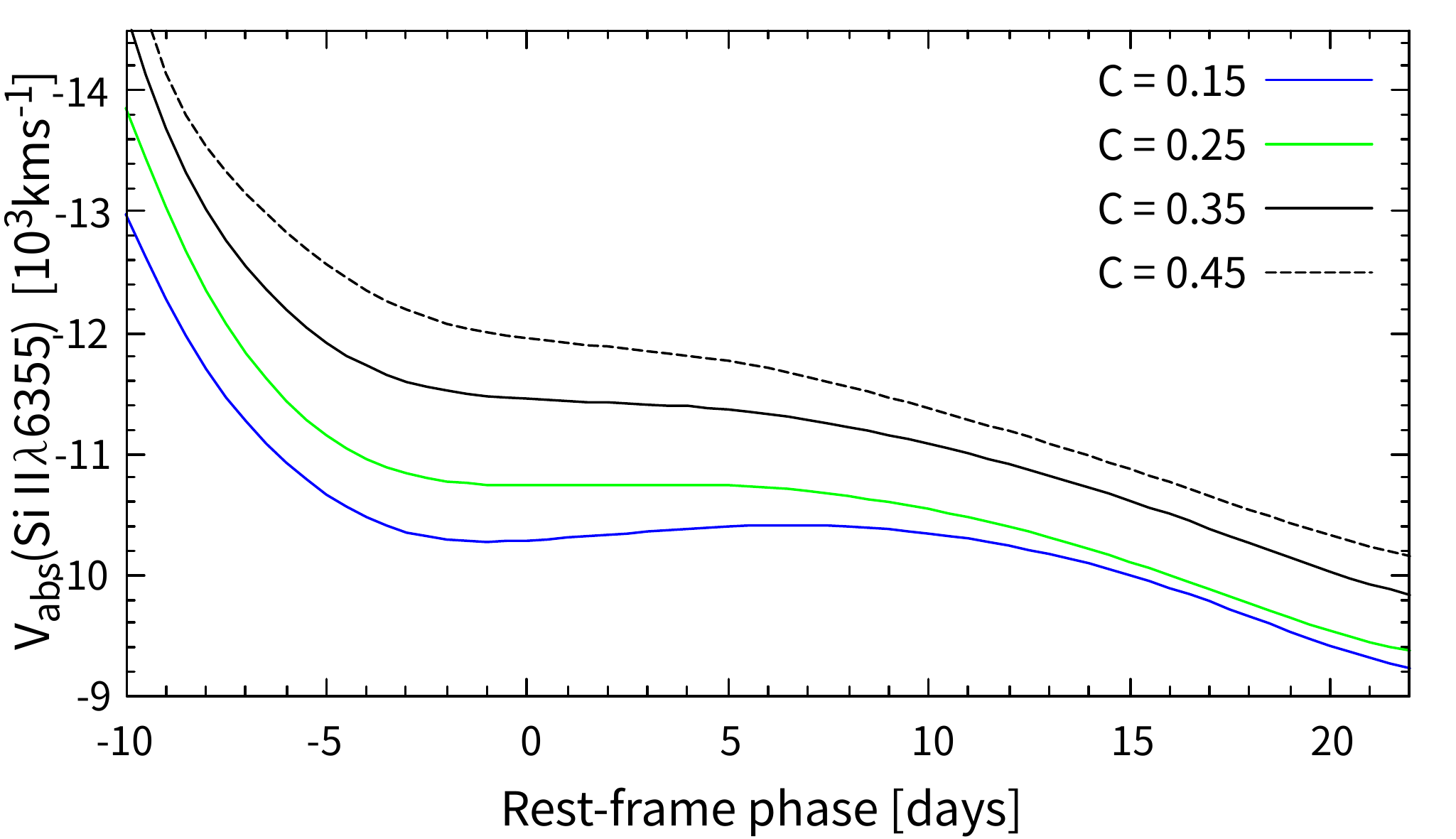}
	\caption{Evolution of the Si II $\lambda$6355 absorption velocity with time for the spectra of different color.}\label{fig:vwitht}
\end{figure}

\begin{figure}[!hbp]
	\includegraphics[width=0.45\textwidth,height=2.3in]{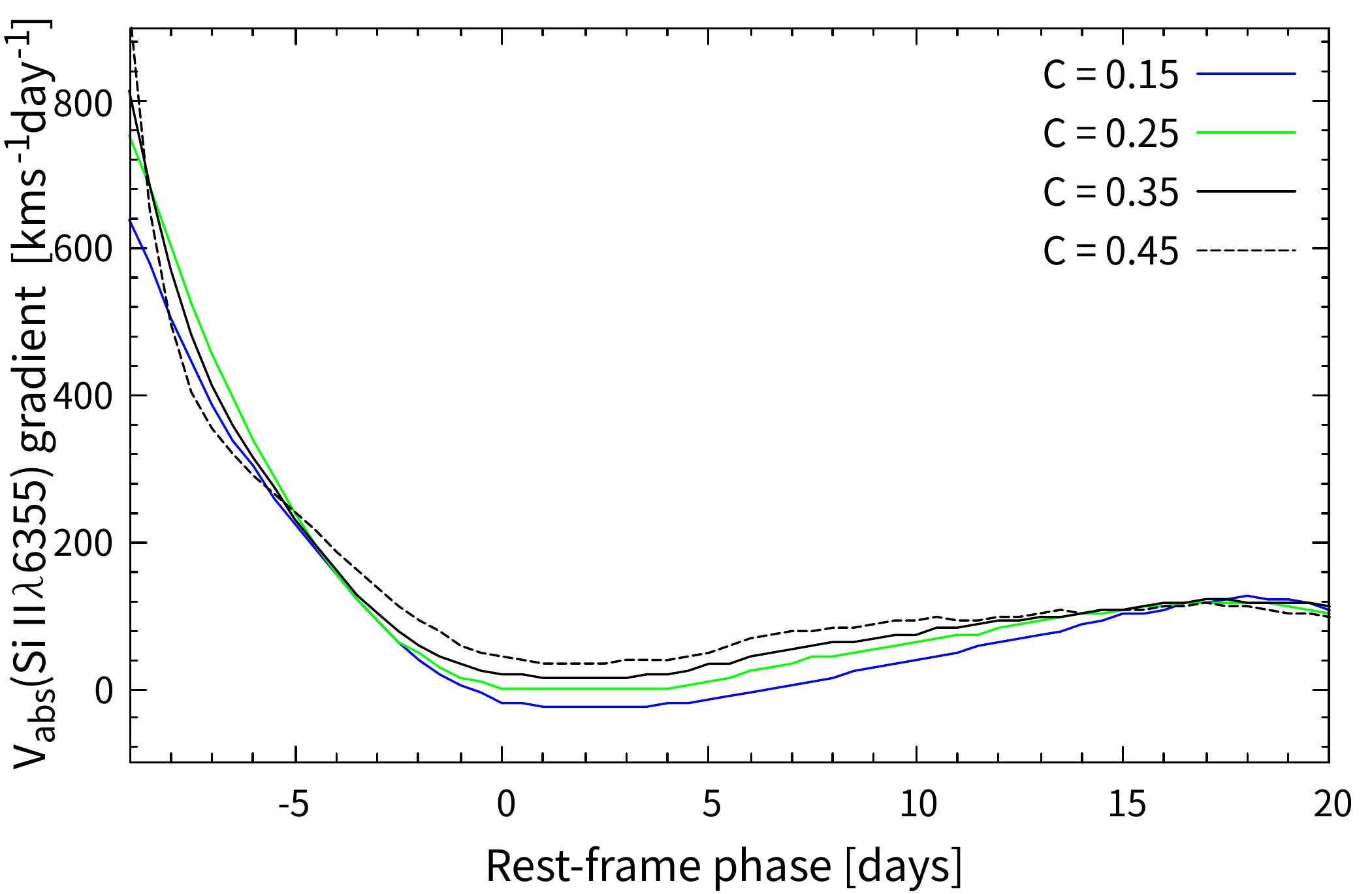}
	\caption{$\partial V_{abs} /\partial t$ as a function of phase.}\label{fig:dvdt}
\end{figure}

The absorption velocity is defined by using the relativistic Doppler formula:
\begin{equation}
\frac{v_{abs}}{c}=\frac{(\lambda_{abs}/\lambda_0)^2-1}{(\lambda_{abs}/\lambda_0)^2+1}\,,
\end{equation}
where $\lambda_0$ is the wavelength of the corresponding transition, and $c$ is the speed of light. The wavelength corresponding to the maximum absorption $\lambda_{abs}(p)$ can be obtained from the trained network, i.e. the value of wavelength  minimizes  $\vec{F}^{ANN}$ at phase $p$, see Fig.\ref{fig:vwitht}. 
And then the changes of $v_{abs}$ can be more efficiently  obtained than those in Ref.\cite{Foley:2011mh}, see Fig.\ref{fig:dvdt}. The evolution is much faster at early times of the supernova's explosion, which is about $v_{abs}\sim 13,000$~km/s at phase $p\sim -8$. Later, the speed decreases to $v_{abs}\sim 11,000$~km/s at the maximum date $p\sim 0$ with range $\sim 2000$~km/s for different colors, and it is nearly a constant during these phases. Finally, it will continue to decrease to $v_{abs}\sim 9,000$~km/s at $p\sim 25$ with range $1,000$~km/s. The speed is almost a constant during $0<p<5$, i,e, $\partial_t v_{abs}\sim0$, while $\partial_t v_{abs}\sim 110$~km/s/day after phase $p\sim10$.

.

\section{Conclusion and future work}\label{sec:conclusion}
We have constructed an SN Ia spectrum evolution model
by training a back-propagation  neural network with the observed nearby spectra from the CfA SN Program. The model can well describe not only the spectrum of SNe Ia with wavelength range from $3500\AA$ to $8000\AA$, but also the light-curve evolution with phase time from $-15$ to $50$ for different colors. We also compare the results with the SALT2 template, and find that the model we constructed by using ANN is almost the same as  the SALT2 template but with much less parameters during the training process.

By taking the advantage of  network method, we calculated  Si II $\lambda6355$ absorption velocity and its gradient in different phase and color, and our results are consistent with the previous works in the literatures.  

It also shows that the Levenberg-Marquardt algorithm for training the network is much faster than others such as the RPROP algorithm. 

With the artificial neural network, the evolution of the universe could be obtained with the trained spectra model, which we will tackle in our future work. Actually, the history of the universe history could be observed without assumption of any cosmological model, for example, the model given in Ref.\cite{Feng:2012gf,Feng:2008kz}. Such a model-independent attempt has been put forward in Ref.\cite{Feng:2016stb}. We will try to achieve this goal using the network method directly.

\acknowledgments
This work is supported by National Science Foundation of China grant Nos.~11105091,~10671128 and~11047138, the Key Project of Chinese Ministry of Education grant No.~211059,``Chen Guang" project supported by Shanghai Municipal Education Commission and Shanghai Education Development Foundation Grant No. 12CG51, and Shanghai Natural Science Foundation, China grant No.~10ZR1422000. This research has made use of the CfA Supernova Archive, which is funded in part by the National Science Foundation through grant AST 0907903.

\end{document}